# The Power of Data Communities

Lucas McCullum (0000-0001-9788-7987)[1,2]
Miguel Ángel Armengol de la Hoz (0000-0002-7012-2973)[3]
Catherine Bielick (0000-0003-1871-3909)[4]
Daniel K. Ebner (0000-0001-6053-5865)[5]
Amelia Fiske (0000-0001-7207-6897)[6]
Jack Gallifant (0000-0003-1306-2334)[7,8]
Judy W. Gichoya (0000-0002-1097-316X)[9]
Rahul Gorijavolu (0000-0002-4386-957X)[10,11]
Nura Izath (0009-0001-6668-2169)[12,*]
Anna E. Premo (0009-0003-3067-3433)[13]
Alice Rangel Teixeira (0000-0001-5034-1223)[14]
Christopher M. Sauer (0000-0002-2388-5919)[11,15,16]
Leo A. Celi (0000-0001-6712-6626)[11,17,18]

*Corresponding Author (nizath@must.ac.ug)
ƗThe middle authors are presented in alphabetical order by last name

[1]UT MD Anderson Cancer Center UTHealth Houston Graduate School of Biomedical Sciences, Houston, USA
[2]Department of Radiation Oncology, The University of Texas MD Anderson Cancer Center, Houston, TX, USA
[3]Big Data Department, PMC, Fundación Progreso y Salud, Seville, Spain
[4]Division of Infectious Diseases and International Health, University of Virginia, Charlottesville, VA, USA
[5]Department of Radiation Oncology, Mayo Clinic, Rochester, MN, USA
[6]Institute of History and Ethics in Medicine, Department of Preclinical Medicine, TUM School of Medicine and Health, Technical University of Munich, Munich, Germany
[7]Artificial Intelligence in Medicine Program, Mass General Brigham, Harvard Medical School, Boston, MA, USA
[8]Department of Radiation Oncology, Brigham and Women's Hospital/Dana-Farber Cancer Institute, Boston, MA, USA
[9]Department of Radiology and Imaging Sciences, Emory University, Atlanta, Georgia, USA
[10]AI for Responsible, Generalizable, and Open Surgical (ARGOS) Research Group, Johns Hopkins University School of Medicine, Baltimore, MD, USA
[11]Laboratory for Computational Physiology, Institute for Medical Engineering and Science, Massachusetts Institute of Technology, Cambridge, MA, USA
[12]Mbarara University of Science and Technology Data Science Research Hub (MUDSReH), Mbarara, Uganda
[13]Department of Humanities, Social, and Political Science, Federal Institute of Technology in Zurich (ETH Zurich), Zurich, Switzerland

[14]Department of Philosophy, Universitat Autónoma de Barcelona, Barcelona, Spain
[15]Department of Hematology and Stem Cell Transplantation, University Hospital Essen, Germany
[16]Institute for Artificial Intelligence in Medicine, University Hospital Essen, Germany
[17]Division of Pulmonary, Critical Care and Sleep Medicine, Beth Israel Deaconess Medical Center, Boston, MA, USA
[18]Department of Biostatistics, Harvard T.H. Chan School of Public Health, Boston, MA, USA

**CRediT Statement**: Conceptualization: LM, LAC; Data curation: LM; Formal Analysis: LM; Visualization: LM; Funding acquisition: CMS, LAC; Methodology: LM, MAAH, DKE, RG, CMS; Project administration: CMS, LAC; Software: LM; Supervision: LAC; Validation: RG; Writing – original draft: LM, MAAH, CB, DKE, AF, JG, JWG, N.A., AEP, ART, RG, CMS, LAC; Writing – review & editing: LM, RG, N.I.

**Acknowledgments:** We acknowledge support by the Open Access Publication Fund of the University of Duisburg-Essen. LM is supported by a National Institutes of Health (NIH) Diversity Supplement (R01CA257814-02S2). JWG is a 2022 Robert Wood Johnson Foundation Harold Amos Medical Faculty Development Program and declares support from Lacuna Fund (#67), Gordon and Betty Moore Foundation, NIH (NIBIB) MIDRC grant under contracts 75N92020C00008 and 75N92020C00021, and NHLBI Award Number R01HL167811. ART is supported by the Spanish Ministry of Science, Innovation and Universities through the State Research Agency under project PID2023-148517NB-I00. CMS is supported by the German Research Foundation co-funded UMEA Clinician Scientist Program (FU356/12-2), the Else Kröner-Fresenius Stiftung (2024_EKEA.130) and the German Federal Ministry of Education and Research (01ZU2403C). LAC is funded by the National Institute of Health through DS-I Africa U54 TW012043-01 and Bridge2AI OT2OD032701, the National Science Foundation through ITEST #2148451, and a grant of the Korea Health Technology R&D Project through the Korea Health Industry Development Institute (KHIDI), funded by the Ministry of Health & Welfare, Republic of Korea (grant number: RS-2024-00403047).

**Competing Interests:** CMS received unrelated consulting fees from Pacmed B.V.

**Data/Code Availability Statement:** All data and code is available at: https://github.com/Lucas-Mc/power-of-data-communities.

# Abstract

Datasets together with active scientific communities prepared to leverage them can contribute to scientific progress and facilitate making research more equitable. In this study we found that MIMIC, despite its limited amount of funding, managed to provide higher impact per dollar spent through accessible data communities. These findings support the notion that making clinical data available empowers innovation which directly addresses clinical concerns and can set new standards for inclusivity.

# 1. Introduction

Restricted access to clinical data undermines research reproducibility and equitable scientific progress. Institutional barriers and privacy concerns limit data sharing, with only 39.4% of studies across 9 disciplines claiming "available upon request" granting access and 19.4% explicitly refusing[1]. This inaccessibility hampers transparency, enabling selective reporting and inflating findings[2]. In healthcare, artificial intelligence (AI) models trained on non-transparent datasets often fail to generalize and introduce biases that disproportionately affect underserved populations[3]. These challenges, as evident in the rapid spread of unverified COVID-19 studies, underscore the need for accessible and collaborative data ecosystems to advance clinical research.

Data communities address these issues by fostering global collaboration with free and de-identified datasets. The Medical Information Mart for Intensive Care (MIMIC) from the Massachusetts Institute of Technology (MIT), hosted on PhysioNet and developed by the Laboratory for Computational Physiology under Dr. Roger Mark, exemplifies this approach. Approximately 10.1% of the researchers utilizing MIMIC data are from low- and middle-income countries and contribute valuable perspectives to the researcher cohort[4]. By lowering barriers to entry, data communities enable broader participation, bridging gaps between well-resourced academic institutions and researchers in underrepresented regions[4–7].

MIMIC's data community fosters innovation in artificial intelligence (AI) and inclusivity through datathons, open-source tools, and transparent credentialing, in contrast to the more controlled access models of datasets like UK Biobank and the National Institutes of Health (NIH). All of Us Research Program, which follow a selective data sharing approach. This study employs bibliometric analysis to compare MIMIC's citation counts, a novel dataset h-index metric, and funding efficiency with those of similar datasets to assess how a data community model influences research productivity, transparency, and equity in healthcare.

# 2. Materials and Methods

We conducted a bibliometric analysis to compare the scientific impact of the MIMIC dataset (versions I-IV) with three similar datasets: UK Biobank, OpenSAFELY, and All of Us. The primary outcome was cumulative citation counts, with secondary outcomes including a novel dataset h-index metric and funding-adjusted citation efficiency, to evaluate how a data community model based on a low-cost initiative influences research productivity, transparency, and equity in healthcare. All data and code is available at: https://github.com/Lucas-Mc/power-of-data-communities.

**Data Sources and Selection**

We included all MIMIC versions (I[8], II[9], III[10], IV[11]) developed by the MIT Laboratory for Computational Physiology (LCP) and hosted on PhysioNet. These data encompass 431,231 hospital stays across 180,733 patients[11]. The other datasets that were selected for comparison include: UK Biobank, created by the Medical Research Council (MRC) and Wellcome Trust and includes genomic and phenotypic data from 500,000 UK participants[12], OpenSAFELY, developed by Ben Goldacre's team at Oxford University with NHS England funding, which covers more than 58 million patient records[13], and All of Us, initiated by the NIH, which collects diverse health data from 633,000 U.S. participants[14]. Other similar datasets, such as AmsterdamUMCdb[15] and HiRID[16], were excluded due to smaller cohorts and later publication date, limiting citation comparability.

**Data Retrieval**

Citation data were manually collected from Google Scholar for peer-reviewed articles published until January 1, 2024, accessed on July 1, 2024. For each dataset, we identified the original publication and accessed its "Cited by" counts. Annual citations were obtained by applying custom year ranges in the interface and recording the counts displayed under the search bar. Citations were aggregated annually using Python 3.7.9 from each dataset's first publication. MIMIC citations were summed across all versions due to its continuous funding and iterative dataset from the same research group. An eight-year period following publication for MIMIC was chosen since after this timepoint, the citations drop significantly since it does not include MIMIC-III anymore. Funding data were manually compiled from publicly available sources: MIMIC (NIH NIBIB/NIGMS/OD grants, 2003-2023); UK Biobank (MRC/Wellcome, GBP converted June 1, 2024); OpenSAFELY (summed NHS grants); and All of Us (NIH budgets). The funding amounts were not adjusted for inflation because the time periods overlap substantially for most datasets (UK Biobank 2015-2024, All of Us 2018-2024, OpenSAFELY 2020-2024), and the magnitude of differences between datasets (ranging from 4-fold to over 800-fold) far exceeds potential inflation effects, which would be approximately 2-3% annually. Secondly, citations were divided by every $1 million in funding to normalize disparities and produce efficiency metrics. Finally, the citations were normalized to the total number of papers citing the respective dataset to account for rate of community impact.

**H-index Calculation for datasets**

H-index is a widely used metric that can be used to measure the productivity and impact of a researcher's publications. In this case, we calculated the h-index of the dataset. The dataset h-index was calculated as the largest *h* where *h* publications citing the dataset's original paper each have at least *h* citations from Google Scholar. Publication counts were normalized by dataset age (years since release) to adjust for time differences.

# 3. Results

We analyzed citations and funding for MIMIC, UK Biobank, OpenSAFELY, and All of Us over their publication periods. The total funding totaled $14,427,192 for MIMIC, $525,546,276 for UK Biobank, $53,715,054 for OpenSAFELY, and $2,160,000,000 for All of Us, as shown in **Table 1**.

**Table 1:** Key access and usage dimensions of the four included datasets.

| Repository | Total Funding (USD) | Restrictions to Open-Access | De-Identified Data | Code Sharing Availability | Community Engagement |
|---|---|---|---|---|---|
| **MIMIC** | $14.4 million | Application, data use agreement | Yes | Yes, https://github.com/MIT-LCP/mimic-code | Datathons |
| **UK Biobank** | $525.5 million | Tiered fees based on data needs | Yes | No | UK Biobank Scientific Conference |
| **OpenSAFELY** | $53.7 million | Limited to pilot users | No (only simulated "dummy" data accessible; code runs remotely) | Yes, https://github.com/opensafely | OpenSAFELY Community Symposium |
| **All of Us** | $2.16 billion | Fees for registration, storage, and computation, US ID verification, training, and data use agreement | Yes, with additional access steps | Yes, https://github.com/all-of-us | Community engagement events |

MIMIC has been available for 27 years (1997-2024), the UK Biobank for nine years (2015-2024), the NIH All of Us for six years (2018-2024), and OpenSAFELY for four years (2020-2024). By shifting to the same time since its release date and adding the citations for each version of MIMIC at the same time since release, the cumulative citations for MIMIC rose from 30 within one year of release to 6,528 by the eighth year as shown in **Figure 1a**. Similarly, UK Biobank citations increased from 18 within one year of release to 7,185 by nine years, All of Us from 1 within one year of release to 852 by six years, and OpenSAFELY from 332 after one year of release to 5,965 by five years. This led to MIMIC, on average, exceeding UK Biobank by a factor of 1.61 and All of Us by 4.44 while trailing OpenSAFELY by a factor of 0.26 over the maximum period each comparative dataset has been released. As shown in **Figure 1b**, adjusting citations by every $1 million in funding (log-10 scale), MIMIC achieved 2 citations per $1 million after one year, rising to 452 by 2024, UK Biobank rose from 0.03 to 14, OpenSAFELY from 0.02 to 12, and All of Us from 0.0005 to 0.4. This led to MIMIC, on average exceeding UK Biobank by 58.81 and All of Us by 664.98 while slightly trailing OpenSAFELY by a factor of 0.96 over the maximum period each comparative dataset has been released. However, MIMIC begins to exceed OpenSAFELY after four years and accelerates while OpenSAFELY slows down.

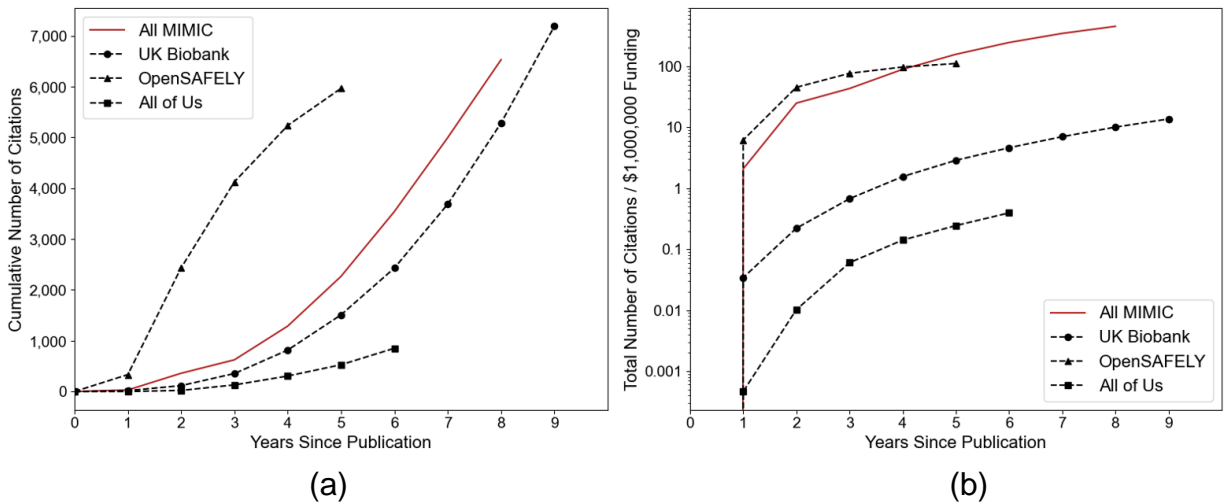

(a)                                                                 (b)

**Figure 1:** The number of cumulative citations for each dataset for each year since publication. a) unadjusted, b) number of citations adjusted per $1 million funding, shown on the log-10 scale. The data in the figures start at 1-year post-publication since no citations are available at year 0 and cannot be represented on the y-axis log plot.

The dataset h-index, defined as the largest $h$ where $h$ publications citing the dataset have $\geq h$ citations, was calculated for each dataset. MIMIC achieved a dataset h-index of 166, UK Biobank with 211, OpenSAFELY with 152, and All of Us with 47 as shown in **Figure 2a**. The dataset h-index, normalized by every $1 million in funding to measure economic efficiency of impact, showed an overall index of 35 for MIMIC, 14 for OpenSAFELY, 4 for UK Biobank, and 0 for All of Us as shown in **Figure 2b**. Quantitatively, on average before adjusting for funding, MIMIC's papers received a factor of 0.67 of the number of citations of UK Biobanks' papers while exceeding OpenSAFELY by a factor of 1.20 and All of Us by a factor of 9.55. However, when adjusting for funding, MIMIC

exceeded the number of citations per $1 million in funding of OpenSAFELY by a factor of 4.47, UK Biobank by a factor of 24.50, and All of Us by a factor of 1430.00.

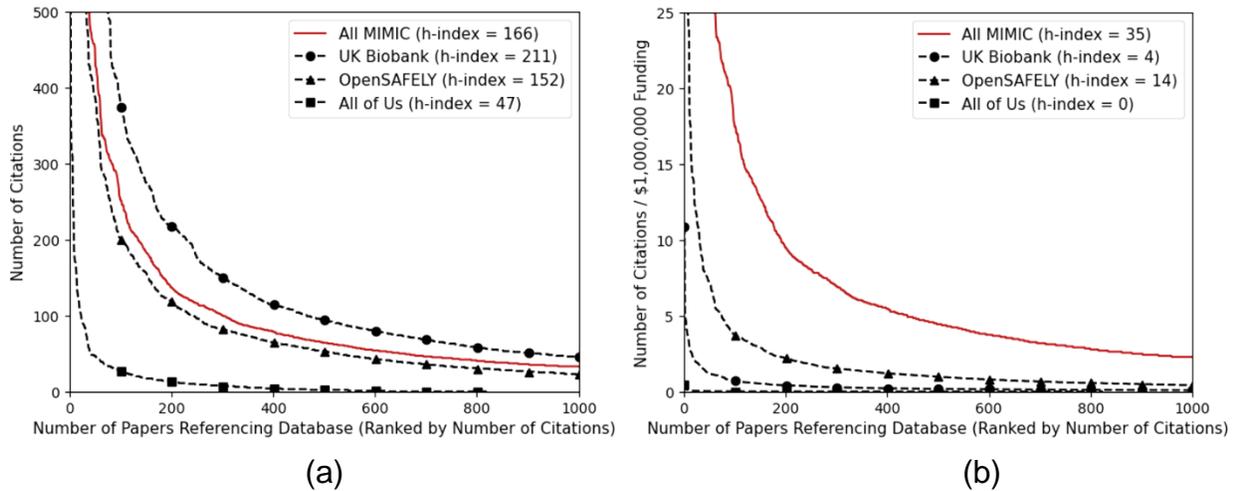

**Figure 2:** Number of papers referencing each dataset ranked by number of citations within each paper. a) unadjusted, b) adjusted per $1 million funding received.

The distribution of citations was analyzed by ranking papers that cited each dataset from highest to lowest citation count. In order to compare datasets with differing numbers of publications, the x-axis was normalized as a percentage of the total number of papers for the dataset papers. Before adjusting for funding, we found that papers citing MIMIC had a citation count that was, on average, a factor of 0.51 less than UK Biobank, a factor of 0.53 less than OpenSAFELY, and a factor of 1.36 greater than of All of Us as shown in **Figure 3a**. However, when adjusting for funding, we found that papers citing MIMIC had a citation count per $1 million in funding that was, on average, a factor of 1.96 greater than OpenSAFELY, a factor of 18.68 greater than UK Biobank, and a factor of 203.98 greater than of All of Us as shown in **Figure 3b**.

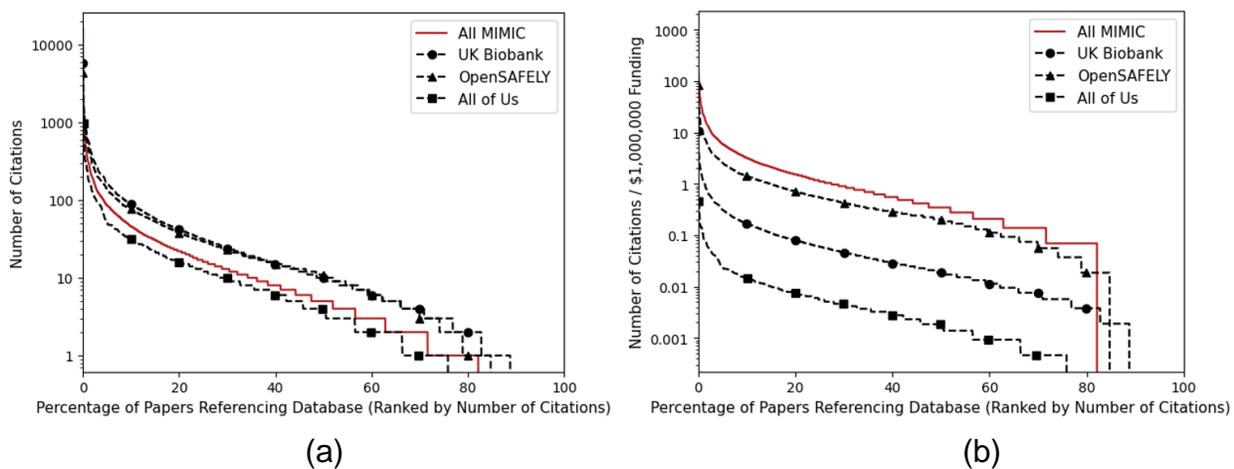

**Figure 3**: Normalized number of papers referencing each dataset ranked by number of citations in each paper. a) unadjusted, b) adjusted per $1 million funding received.

# 4. Discussion

Datasets like MIMIC hold immense potential but require intentional efforts to broaden inclusivity, accessibility and engagement. Our analysis of community engagement characteristics (**Table 1**) suggests that active community cultivation may be a key differentiator in research productivity and inclusivity outcomes. The MIT Critical Data community (https://criticaldata.mit.edu/) harnesses MIMIC as a resource in advancing health equity and is guided by five core values: (1) rigorous and innovative research, (2) multi-level and accessible teaching, (3) building networks of primary stakeholders, (4) reimagining legacy systems of power, and (5) advocacy for epistemic humility and health equity. Since 2014 and driven by these values, MIT Critical Data has hosted 46 datathons across 21 countries, providing platforms that spark rich inter-disciplinary collaborations and fresh approaches to combat bias, ensuring AI systems reflects the priorities of diverse communities.

This work has yielded over 2,000 publications, with approximately 10.1% from low- and middle-income countries (LMIC), compared to 6.2% for UK Biobank[4]. Similarly, minority-serving institutions (MSIs) in the U.S. accounted for 8.6% of MIMIC-related publications, versus 5.6% in more closed datasets. The MIT Critical Data community attributes MIMIC's enhanced global participation to deliberate community cultivation efforts that distinguish it from more closed datasets (which charge fees), restricted datasets (which require extensive credentialing), and passively open datasets (which are freely available but lack active community engagement infrastructure).

Closed systems often limit participation to well-resourced groups, which can narrow research scope and limit engagement greatly[17]. In contrast, open data communities provide the context necessary for enabling equitable access to data, networking opportunities, targeted training and a collaborative culture, to support equitable engagement for underfunded institutions and low-resource communities. Furthermore, they offer capacity building, equipping participants with technical skills, ethical practices, and contextual understanding for responsible use and interpretation of data. Institutions in LMICs can leverage these benefits through mentorship, shared analytical tools, and cross-disciplinary exchange, allowing multidisciplinary participants to co-create solutions tailored to priorities of diverse communities, in addition to academic productivity. Additionally, open data communities have the potential to build public trust in science by encouraging transparency and reproducibility to ensure data-driven policies are accepted and implemented. This intentional and inclusive engagement is key to facilitating collaborations that reduce global research disparities, drive innovation through diverse problem-solution fit approaches, and contribute to real world health outcomes using research data.

To actualize the value of open datasets, we must rethink our approach to include an ecosystem fostering sustained investment for research at the individual, institutional, federal, and global level. As individual researchers, we must share resources including code, tools, mentor newcomers, curate documents for easy accessibility and alignment with specific research needs and reach out to researchers at institutions in other parts of the world, fostering diverse contributions[18]. Institutions should invest in accessible platforms or develop user-friendly data platforms and provide targeted training to lower barriers of engagement for underrepresented or under-resourced groups. At the federal level, we must continue to incentivize meaningful data sharing through policies and efforts such as the NIH Data Sharing Index (S-Index). Globally, researchers must work together to establish channels of collaboration and perspective sharing[19], such as encouraging open access to research data, providing guidelines for data sharing policies, frames for data management and reuse. These multi-level efforts build communities that ensure open data aligns with open knowledge exchange, hence transforming open data into equitable and innovative science that optimizes our investments into research[17].

This study has several limitations. First, our analysis is constrained by the heterogeneity of datasets, which vary in data structures, coding systems, and access protocols. These differences limit the ability to conduct standardized comparisons across repositories and may result in analytical inconsistencies. Second, our assessment relies on bibliometric metrics such as publication counts and citation rates, which while valuable, may not fully capture the broader societal impact of open data, including shifts in clinical practice, policy influence, or community engagement[20]. For example, OpenSAFELY's publication was focused on the clinical question of COVID-19 which may have inflated its citations compared to the more technical papers of the other datasets. However, OpenSAFELY recently released a technical description of its dataset[21] which may be used for comparison in future studies. While this paper focuses on health-related data repositories, our findings may not be generalizable to open data initiatives in other fields, limiting broader cross-sectoral insights. Third, comparing MIMIC to the other datasets presents specific challenges since multiple versions have been released over time, each with their own corresponding publications, which may potentially introduce version related bias in the results.

# 5. Conclusion

In sum, open data sets the stage to advance research. The MIT Critical Data community's efforts demonstrate how deliberate engagement amplifies the impact of accessible datasets and cultivates global collaboration and equitable innovation. Beyond citation metrics, open data empowers a broader range of researchers to address pressing healthcare challenges, hopefully helping to develop solutions that reflect global needs. The path forward lies in building robust infrastructure for collaboration, strengthening collaborative networks, and supporting policies that prioritize meaningful data sharing. By embracing these principles, open data is one step towards dismantling inequities, enabling a research ecosystem where all voices have the possibility to contribute to impactful science. This vision demands collective action to ensure open data's potential is fully realized, advancing health equity worldwide.